\documentclass[pra,twocolumn,showpacs]{revtex4}
\usepackage{graphicx}
\usepackage{amsmath}
\usepackage{amssymb}
\usepackage{dcolumn}
\usepackage{float}
\usepackage{bm}
\usepackage[breaklinks=true,colorlinks,citecolor=blue,linkcolor=blue,urlcolor=blue]{hyperref}

\DeclareMathAlphabet{\bi}{OML}{cmm}{b}{it}

\def\be{\begin{equation}}
\def\ee{\end{equation}}
\def\bearr{\begin{eqnarray}}
\def\eearr{\end{eqnarray}}
\def\la{\langle}
\def\ra{\rangle}

\def\bs{\boldsymbol}

\begin{document}
\title{Zitterbewegung of a
heavy hole in presence of spin-orbit interactions}
\bigskip
\author{Tutul Biswas$^{1,2}$} 
\email{tutulb@iitk.ac.in, tbtutulm53@gmail.com}
\author{Sandip Chowdhury$^1$}
\author{Tarun Kanti Ghosh$^1$}
\normalsize
\affiliation
{$^1$Department of Physics, Indian Institute of Technology-Kanpur,
Kanpur-208 016, India\\
$^2$Department of Physics, Vivekananda Mahavidyalaya-Burdwan, Sripally-713 103, West Bengal, India}
\date{\today}

\begin{abstract}
We study the $zitterbewegung$ of a heavy hole in presence of both cubic Rashba and cubic Dresselhaus 
spin-orbit interactions. On contrary to the electronic case, $zitterbewegung$ does 
not vanish for equal strength of Rashba and Dresselhaus spin-orbit interaction. This non-vanishing of $zitterbewegung$
is associated with the Berry phase. Due to the presence of the spin-orbit
coupling the spin associated with the heavy hole precesses about an effective magnetic field. This spin precession
produces a transverse spin-orbit force which also generates an electric voltage associated with $zitterbewegung$.
We have estimated the magnitude of this voltage for a possible experimental detection of $zitterbewegung$.

\end{abstract}
 \pacs{71.70.Ej, 73.21.Fg, 03.65.-w }

\maketitle


\section{Introduction}

According to Schr\"{o}dinger\cite{Schro}, the interference between two branches of a free 
Dirac spectrum induces a quivering quantum motion, usually known as 
{\it zitterbewegung} (ZB). In principle, ZB is not a pure relativistic phenomenon.
In recent years, it has been shown that ZB could exist in a plethora of non-relativistic physical
systems\cite{zbgen} including narrow-gap semiconductors\cite{zawadki}, spin-orbit coupled low-dimensional
systems\cite{john,zb2d1,zb2d2,zb2d3,zb2d4,zb2dH,zb2d5}, graphene\cite{zbgrph1,zbgrph2,zbgrph3,zbgrph4,zbgrph5},
carbon nanotube\cite{cnt}, topological insulator\cite{zbtopo}, superconductors\cite{zbsup}, 
sonic crystal\cite{zbsonic}, photonic crystal\cite{zbphoton},
optical superlattice\cite{zbopp}, Bose-Einstein condensates\cite{zbbec1,zbbec2,zbbec3,zbbec4}, 
dichalcogenide materials like MoS$_2$\cite{zbMos2,zbMos22} etc.

The existence of spin-orbit interaction (SOI)\cite{spin1} in systems like two-dimensional 
electron/hole gas (2DEG/2DHG) is an interesting topic of contemporary research
due to the promising spintronic applications\cite{spindv1,spindv2,spindv3,spindv4,spindv5}. In addition, these systems exhibit various
fundamental physical phenomena such as zero field spin-splitting\cite{zeroFS}, spin Hall effect\cite{SHE1,SHE2,SHE3,SHE4,SHE5},
 persistent spin helix\cite{helix1,helix2} etc. The absence of structural and bulk inversion symmetries
in semiconductor heterostructures induce Rashba\cite{rashba1, rashba2} and Dresselhaus\cite{dress1} SOIs, respectively.  
The functional dependence of both SOIs on momentum in 2DEG and 2DHG
are different. In the case of a 2DEG, they are linear in momentum whereas 
for the 2DHG both SOIs are cubic in momentum. Systems with cubic SOIs are 
particularly important as large spin Hall conductivity can be achieved in those systems\cite{sheAd1,sheAd2,sheAd3}. Apart from
p-doped semiconductor heterostructures, the existence of cubic SOIs have been confirmed
experimentally in a 2DEG at the surface of SrTiO$_3$\cite{cubicSr} and in a 2DHG in a strained-Ge/SiGe
quantum well\cite{cubicGe}.

The problem of ZB of an electronic wave packet in a 2DEG including both Rashba and
Dresselhaus SOIs is well studied. To the best of our knowledge, a less effort has
been devoted in searching ZB in a spin-orbit coupled 2DHG.
In this work, we study the ZB of the center of a Gaussian wave packet
in a 2DHG in presence of both cubic Rashba and Dresselhaus SOIs. We find that
the ZB exhibits transient behavior. The amplitude of ZB is related with the
Berry connection. Unlike the case of a 2DEG, the ZB survives for equal strength
of the Rashba and Dresselhaus SOI. We relate this non-vanishing of ZB with the behavior of 
the associated Berry phase. The spin associated with the heavy hole precesses about a SOI induced
effective magnetic field. A transverse component of the spin-orbit force is generated
as a result of this spin precession. Using this spin-orbit force picture, we have also
estimated the magnitude of an induced electric voltage associated with ZB.

The rest of the paper is organized as follows. In section II, we briefly discuss
the preliminary informations about the physical system. In section III, we consider
the temporal evolution of an initial Gaussian wave packet and discuss various features of the ZB
in position and velocity. The time evolution of the spin and the spin-orbit force picture for ZB
are presented in section IV. We provide asymptotic expressions for ZB and possible detection scheme
in section V. The main results are summarized in section VI.

\section{Physical system}
Let us start with a brief description of the model Hamiltonian associated
with the spin-orbit coupled 2DHG. The complicated dynamics of holes at the top most valence band 
of a III-V semiconductor is characterized by the $4\times4$ Luttinger Hamiltonian\cite{lutt1}. 
However, strong confinement in a p-doped III-V quantum well essentially leads 
to a large splitting between the heavy hole (HH) state ($\vert 3/2,\pm3/2\rangle$) and 
the light hole (LH) state ($\vert 3/2,\pm1/2\rangle$). At low temperature, it is assumed that only the HH 
states are occupied when the density is low enough. Now, one can proceed without considering
the LH states since a significant contribution to the transport properties near the Fermi energy comes from the 
HH states. Hence, it is possible to obtain an effective $2\times2$ Rashba Hamiltonian
by projecting the $4\times4$ Luttinger Hamiltonian onto the HH states\cite{sheAd1}. Additionally
the Dresselhaus SOI originates due to the bulk inversion asymmetry of the host crystal\cite{dressH}.

Now, the single particle dynamics of a HH is governed by the following Hamiltonian
\begin{eqnarray}\label{ch3Ham}
H &=& \frac{{\bf p}^2}{2m^\ast} +\frac{i\alpha}{2\hbar^3}\Big(\sigma_{+}p_{-}^3-\sigma_{-}p_{+}^3\Big)\nonumber\\
&-&\frac{\beta}{2\hbar^3}\Big(\sigma_+p_-p_+p_-+\sigma_-p_+p_-p_+\Big),
\end{eqnarray}
where $\sigma_{\pm}$ = $\sigma_{x}\pm i\sigma_{y}$ with $\sigma_i$'s as
the usual Pauli spin matrices and $p_\pm = p_x\pm ip_y$
with ${\bf p}$ is the momentum, $m^\ast$ is the effective mass 
of the heavy hole and $\alpha$ ($\beta$) is the strength of the 
Rashba (Dresselhaus) SOI. Note that $\alpha$ can be tuned by an
external gate voltage but $\beta$ is a fixed material dependent
quantity. Here, Pauli matrices represent an effective pseudo-spin with 
spin projection $\pm3/2$ along the growth direction of the quantum well.

The eigenvalues and eigenstates of the Hamiltonian are respectively given by\cite{Tot2D},
\begin{eqnarray}\label{ch3eigen}
\epsilon_{\bf k}^\lambda=\frac{\hbar^2k^2}{2m^\ast}+\lambda\Delta_\theta k^3,
\end{eqnarray}

and
\begin{eqnarray} \label{ch3eigstate}
\psi_{\bf k}^\lambda({\bf r})=\frac {e^{i{\bf k} \cdot r}} {2\sqrt{2}\pi}\begin{pmatrix} 1
\\\lambda e^{i(2\theta-\phi_k)} \end{pmatrix}, 
\end{eqnarray}
where $\lambda=\pm$, $\Delta_\theta=\sqrt{\alpha^2+\beta^2-2\alpha \beta \sin{2\theta}}$ 
with $ \theta = \tan^{-1}(k_y/k_x)$ and
$\phi_k =\tan^{-1}[(\alpha k_x-\beta k_y)/(\alpha k_y-\beta k_x)]$. 
Note that the energy spectrum (Eq.~\eqref{ch3eigen}) is highly anisotropic 
described by the factor $\Delta_\theta$.
The spin splitting between the HH branches at a given wave vector can be 
obtained as $\epsilon_{\bf k}^+$ - $\epsilon_{\bf k}^-$=$2\Delta_\theta k^3$.

\section{Time evolution}
Here, we seek to see the time evolution of an 
initial wave packet representing a HH in the presence of the SOIs. 
Now, applying the time evolution
operator $ U(t) = e^{-iHt/\hbar} $ on the initial Gaussian wave 
packet polarized along $+z$ axis
\begin{eqnarray}\label{initial}
\Psi\big({\bf r},0 \big)=\frac{1}{2\pi}\int\,d^2k\, a({\bf k}, 0)
e^{i\bf k \cdot r}
\left(\begin{array}{c}
1 \\0 \end{array}\right),
\end{eqnarray}
we find the wave packet at a later time $t$ as

\begin{eqnarray} \label{time_wp}
\Psi({\bf r}, t)&=&\frac{1}{2\pi}\int\ d^2k\, a({\bf k}, 0)e^{i\bf k \cdot r}
e^{-i\frac{\hbar k^2}{2m^\ast}t} \nonumber\\
&\times&\bigg\{\cos(\omega_{\bf k}t)\left(\begin{array}{c}
1 \\0 \end{array}\right)+g_k^\ast\sin(\omega_{\bf k}t)\left(\begin{array}{c}
0 \\1 \end{array}\right)\bigg\},
\end{eqnarray}
where $a({\bf k},0) = d/(\sqrt{\pi}) e^{-\frac{1}{2} d^2\big({\bf k-k}_0 \big)^2}$
with $d$ (${\bf k}_0$) is the width (wave vector) of the initial wave packet,
$g_k=ie^{-i(2\theta-\phi_k)}$ and $\omega_{\bf k}=(\epsilon_{\bf k}^+
-\epsilon_{\bf k}^-)/2\hbar$. However, the wave packet was initially polarized along 
$+z$ direction but it acquires a component along $-z$ direction as time goes on.

By taking the inverse Fourier transformation of Eq.~\eqref{time_wp}, we obtain 
the following expression for the wave packet in the momentum space at time $t$
\begin{eqnarray}\label{ch3fr}
\Phi({\bf k},t)=a({\bf k}, 0)e^{-i\frac{\hbar k^2}{2m^\ast}t}\left(\begin{array}{c}
\cos(\omega_{\bf k}t)\\ g_k^\ast\sin(\omega_{\bf k}t)\end{array}\right).
\end{eqnarray}

\subsection{Calculation of the expectation values}
Here, we would like to calculate the expectation values of position
and velocity operators explicitly.
The expectation value of any physical operator $\hat{O}$ is defined 
via $\la \hat{O}\ra=\int d^2k \Phi^\dagger({\bf k},t)\hat{O}\Phi({\bf k},t)$.
After a lengthy calculation, one can obtain the average value of the position operator as
\begin{eqnarray}\label{ch3Pos} 
\la {\bf r}(t)\ra&=&\la {\bf r}(0)\ra +\frac{\hbar \la{\bf k}(0)\ra}{m^\ast}t
-\frac{1}{2}\int d^2k \left|a({\bf k,0})\right|^2\nonumber\\
&\times&\Big({\bf \nabla}_{\bf k}(2\theta-\phi_k)\Big)\Big\{1-\cos(2\omega_{\bf k}t)\Big\},
\end{eqnarray}
where $\la {\bf r}(0)\ra$ and $\la {\bf k}(0)\ra$ can be determined 
from the initial condition. The last oscillatory term in Eq.~\eqref{ch3Pos}
represents the phenomenon $zitterbewegung$, the frequency of which
is governed by the spin-split energy differences
i.e $\omega_{\bf k}=(\epsilon_{\bf k}^+$ - $\epsilon_{\bf k}^-)/(2\hbar)$.

Now, setting $\la {\bf r}(0)\ra\equiv0$ and $\la {\bf k}(0)\ra\equiv {\bf k_0}=k_0\hat{y}$, we 
obtain the following expressions for the expectation values of the position operator
\begin{eqnarray} \label{Ch3posX}
\la x(t) \ra& =&
\frac{d}{\pi}e^{-a_0^2}\int^{2\pi}_0\sin\theta d\theta
\int_0^\infty dq\, e^{-q^2+2a_0q\sin\theta}\nonumber\\ 
&\times&\bigg(1+\frac{1-\eta^2}{2Q^2}\bigg)
\big[1-\cos(\Omega_qt)\big],
\end{eqnarray}
and
\begin{eqnarray} \label{Ch3posY}
\la y (t) \ra  &=&\frac{\hbar k_0}{m^\ast}t
-\frac{d}{\pi}e^{-a_0^2}\int^{2\pi}_0\cos\theta d\theta
\int_0^\infty dq\, e^{-q^2+2a_0q\sin\theta}\nonumber\\
&\times&\bigg(1+\frac{1-\eta^2}{2Q^2}\bigg)
\big[1-\cos(\Omega_qt)\big],
\end{eqnarray}
where $q=kd$, $a_0=k_0d$, $Q =\sqrt{1+\eta^2-2\eta \sin(2\theta)}$ with
$\eta=\beta/\alpha$, and $\Omega_q=2\alpha q^3Q/(\hbar d^3)$. 
Note that when $\beta=0$, there will be no oscillatory term in Eq.~\eqref{Ch3posY} 
since the integral $\int_0^{2\pi}e^{2a_0q\sin\theta}\cos\theta d\theta$
is exactly zero. However, for a finite $\beta$, the angular integration can not
be done analytically. Nevertheless, one can show numerically that the 
second term in Eq.~\eqref{Ch3posY} is negligibly small compared to the first term.
In this way, one can say that
the ZB essentially occurs in a direction perpendicular to the initial wave vector. Henceforth,
we will consider only the $x$ component of the observables.

The $x$ component of the velocity operator can be obtained using the 
Heisenberg equation $ v_x=[x,H]/(i\hbar)$ as
\begin{eqnarray}
v_x&=&\frac{p_x}{m^\ast}\sigma_0+\frac{3i\alpha}{2\hbar^3}
\Big(\sigma_{+}p_{-}^2-\sigma_{-}p_{+}^2\Big)\nonumber\\
&-&\frac{\beta}{2\hbar^3}\Big\{\sigma_{+}(2p^2+p_{-}^2)
+\sigma_{-}(2p^2+p_{+}^2)\Big\}.
\end{eqnarray}

After a straightforward calculation, we find the expectation value 
of $v_x$ as 
\begin{eqnarray} \label{ch3velX}
\la v_x(t)\ra&=& 
\frac{v_r}{\pi}e^{-a_0^2}\int^{2\pi}_0  d\theta
\int_0^\infty dq\,e^{-q^2 + 2a_0 q\sin\theta}\sin\theta g(\theta)
\nonumber\\
& \times &\frac{q^3}{Q}
\sin(\Omega_qt),
\end{eqnarray}
where $v_r=\alpha/(\hbar d^2)$ and
$g(\theta)=\eta^2+3-4\eta\sin(2\theta)$.

\subsection{Non-vanishing of zitterbewegung for $ \alpha=\beta$}
A close inspection of Eqs.~\eqref{Ch3posX} and \eqref{ch3velX} reveals
that the ZB appearing in either position or velocity does not vanish for
$\alpha=\beta$. This result strongly contradicts the electronic case 
in which it was argued\cite{zb2d2} that ZB would vanish for equal strength of Rashba and
Dresselhaus SOI because of the existence of an additional
conserved quantity $\sigma_x-\sigma_y$. But for the case of a $2$DHG no such conserved quantity
exists. However, here we try to relate the non-vanishing of ZB
with the Berry phase associated with the energy spectrum.
The Berry connection is given by 
${\bf A_k}=\la \psi_{{\bf k}s}^{\lambda}
\vert i \frac{\partial}{\partial {\bf k}}\vert \psi_{{\bf k}s}^{\lambda} \ra$,
where $\psi_{{\bf k}s}^{\lambda}$ denotes the spinor part of the wave functions
given in Eq.~\eqref{ch3eigstate}.
An explicit calculation will yield 
\begin{eqnarray}\label{ch3BerryC}
{\bf A_k}=\frac{1}{k^2}\bigg(2+\frac{\alpha^2-\beta^2}{\Delta_\theta^2}\bigg)
(k_x\hat{y}-k_y\hat{x}).
\end{eqnarray}

The corresponding Berry phase can be obtained by
taking the line integral of the Berry connection as
$\gamma=\oint {\bf A_k}\cdot {\bf dk}$.
Using the residue theorem of complex integration 
we find the Berry phase as
\begin{eqnarray}\label{ch3BerryP}
\gamma=2\pi+\pi\frac{\alpha^2-\beta^2}{\vert \alpha^2-\beta^2\vert}.
\end{eqnarray}
 
Eq.~\eqref{ch3BerryP} clearly shows that the Berry phase does 
not vanish for the case of $\alpha=\beta$. It has been 
found\cite{BerryP} for the 2DEG case that the Berry phase becomes zero 
when the strengths of Rashba and Dresselhaus SOI are equal.

\begin{figure}[h]
\begin{minipage}[b]{\linewidth}
\centering
 \includegraphics[width=7.5cm, height=2.4cm]{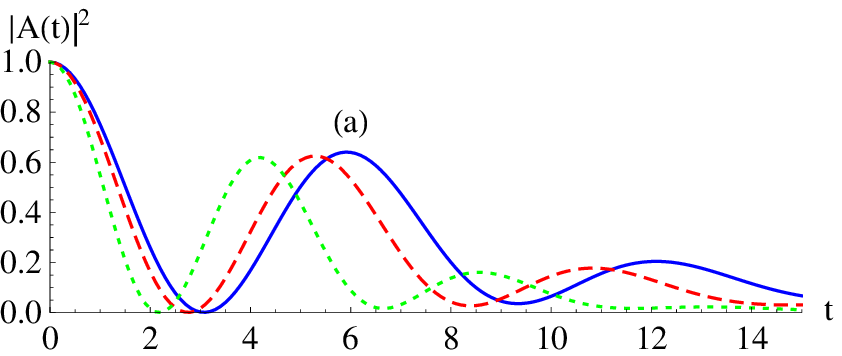}
\end{minipage}

\begin{minipage}[b]{\linewidth}
\centering
\includegraphics[width=7.5cm, height=2.4cm]{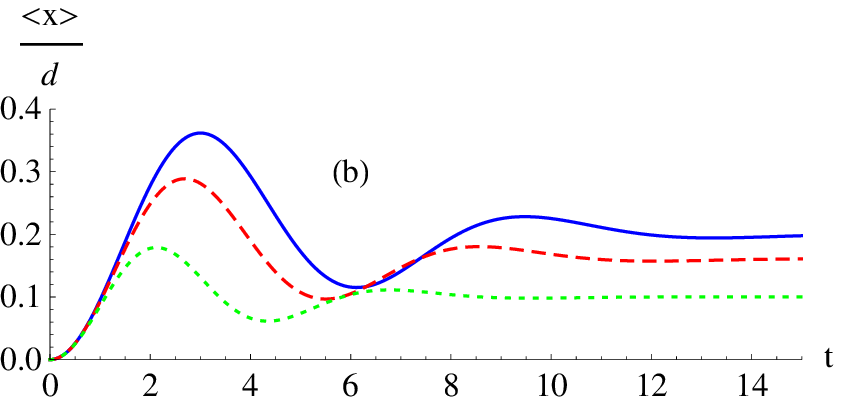}
\end{minipage}

\begin{minipage}[b]{\linewidth}
\centering
\includegraphics[width=7.5cm, height=2.4cm]{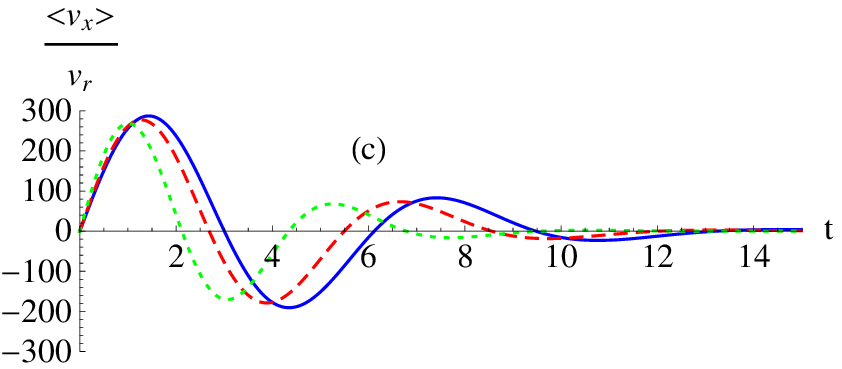}
\end{minipage}
\caption{Plots of (a) square of the autocorrelation function,
(b) position, and (c) velocity expectation values for $a_0=10$. Solid, dashed, and 
dotted line represent $\eta=0$, $0.5$, and $1$, respectively. Here, $t$ is considered in units of $10^{-3}\hbar d^3/(2\alpha)$.}
\label{ch3fig1}
\end{figure}

\subsection{Transient zitterbewegung in position and velocity}
In Fig. \ref{ch3fig1} we have shown the time variation of the
square of the autocorrelation function ($\vert A(t)\vert^2$), expectation values of
position $\la x(t)\ra$ and velocity $\la v_x(t)\ra$ operators for a fixed $a_0=10$ and
different $\eta=0$, $0.5$, and $1$. The autocorrelation 
function mainly measures the overlap of the time evolved wave
packet with its initial counterpart and is defined as 
$A(t)=\la \Psi({\bf r},t)\vert\Psi({\bf r},0)\ra=
\int d^2k\, \Phi^\dagger({\bf k},t)\Phi({\bf k},0)$. Now using
Eq.~\eqref{ch3fr}, it is straightforward to find $A(t)$ in the
following form
\begin{eqnarray}\label{ch3auto}
A(t)&=&\frac{1}{\pi}e^{-a_0^2}\int d\theta\, 
dq\, q\, e^{-q^2+2a_0q\sin\theta}\, e^{i\frac{\hbar q^2t}{2m^\ast d^2}}\cos\Big(\frac{\Omega_qt}{2}\Big)\nonumber\\
\end{eqnarray}

Note that for a finite $\alpha\neq0$, $\beta\neq0$, the integrals
over $q$ and $\theta$ appearing in Eqs.~\eqref{Ch3posX}, \eqref{ch3velX},
and \eqref{ch3auto} can not be done exactly and hence numerical
treatment has to be implemented to see the explicit time dependence of
$\vert A(t)\vert^2$, $\la x(t)\ra$, and $ \la v_x(t)\ra$. From Fig. \ref{ch3fig1}(b)
and Fig. \ref{ch3fig1}(c) it is evident that the ZB, appearing in position and
velocity, shows transient oscillations. The time scale associated with ZB can not be
extracted analytically. However, the transient character of $\la x(t)\ra$ and $ \la v_x(t)\ra$ may be explained 
from the behavior of the autocorrelation function which is shown in Fig. \ref{ch3fig1}(a).
$\vert A(t)\vert^2$ exhibits damped oscillatory character. Fig. \ref{ch3fig1} clearly shows that the oscillations
associated with ZB stop when the autocorrelation function nearly dies out.
Now the effect of an enhancement of $\beta$ on ZB is twofold. First, it introduces 
a phase in the oscillations and second, the amplitude of ZB decreases with the increase of $\beta$ as 
evident from Fig. \ref{ch3fig1}(b).

\begin{figure}[h!]
\begin{minipage}[b]{\linewidth}
\centering
 \includegraphics[width=7.5cm, height=2.6cm]{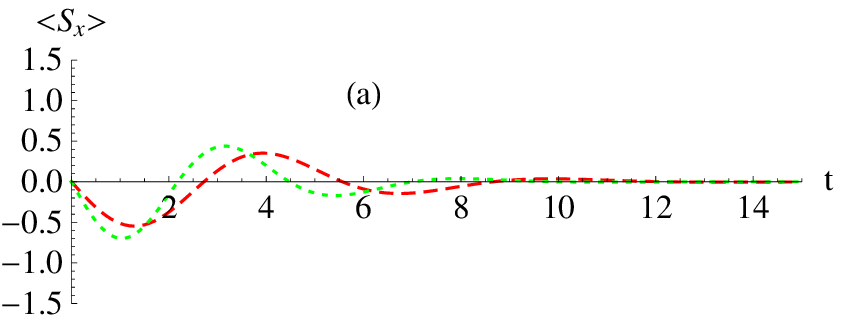}
\end{minipage}

\begin{minipage}[b]{\linewidth}
\centering
\includegraphics[width=7.5cm, height=2.6cm]{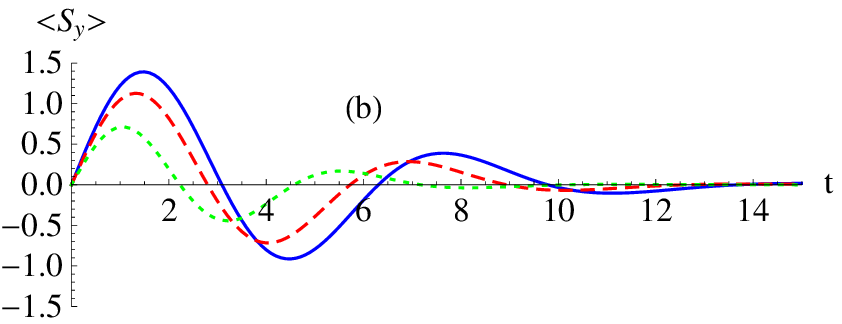}
\end{minipage}

\begin{minipage}[b]{\linewidth}
\centering
\includegraphics[width=7.5cm, height=2.6cm]{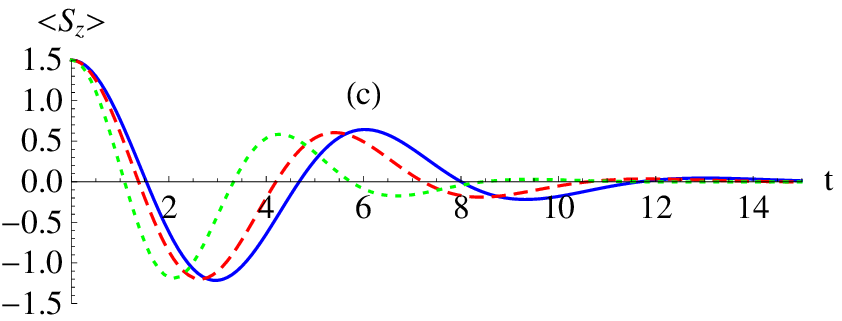}
\end{minipage}
\caption{Plots of the expectation values of the spin operators for $a_0=10$. Solid, dashed, and 
dotted line represent $\eta=0$, $0.5$, and $1$, respectively. Here, $t$ is considered in units of $10^{-3}\hbar d^3/(2\alpha)$.}
\label{ch3fig4}
\end{figure}

\section{Spin dynamics and spin-orbit force}
Here, we are interested to study the time dependence of the
expectation values of the effective spin operator and the spin-orbit force.
Due to the existence of SOIs, the spin ${\bf S}=(3/2)\hbar {\bs \sigma}$ associated with the
HH precesses about an effective magnetic field ${\bf \Omega}$ in accordance with the following equation
\begin{eqnarray}\label{spin_precs}
\frac{d{\bf S}}{dt}={\bf \Omega}\times{\bf S}.
\end{eqnarray}
The components of $\bf{\Omega}$ can be obtained as
\begin{eqnarray}
\Omega_x=\frac{2}{\hbar^4}[\alpha p_y(3p_x^2-p_y^2)-\beta p^2p_x]\nonumber
\end{eqnarray}

and
\begin{eqnarray}
\Omega_y=\frac{2}{\hbar^4}[\alpha p_x(3p_y^2-p_x^2)-\beta p^2p_y].\nonumber
\end{eqnarray}

The general solution of Eq.~\eqref{spin_precs} is given by
\begin{eqnarray}
{\bf S}&=&(\hat{n}\cdot{\bf S}_0)\hat{n}+[{\bf S}_0-(\hat{n}\cdot{\bf S}_0)\hat{n}]
\cos(\Omega t)\nonumber\\
&+&(\hat{n}\times{\bf S}_0)\sin(\Omega t),
\end{eqnarray}
where $\Omega=\sqrt{\Omega_x^2+\Omega_y^2}$, ${\bf S}_0$ is
determined from the initial condition, and the unit vector $\hat{n}$ is
defined as $\hat{n}={\bf \Omega}/\Omega$. Without any loss of generality, 
we choose the initial condition such that $\hat{n}\cdot{\bf S}_0=0$ or
equivalently, ${\bf S}_0=S_z\hat{z}$ since $\hat{n}$ lies in the $xy$ plane.
So, one can obtain the components of the effective spin operator at a later time $t$
as $S_x(t)=(\Omega_y/\Omega)S_z\sin(\Omega t)$, $S_y(t)=-(\Omega_x/\Omega)S_z\sin(\Omega t)$,
and $S_z(t)=S_z\cos(\Omega t)$. We find the expectation values of the spin components 
at time $t$ as

\begin{eqnarray}\label{SpnExpt1}
\left(\begin{array}{c}
\la S_x(t)\ra \\\la S_y(t)\ra \end{array}\right)
&=&-\frac{3\hbar}{2\pi}
e^{-a_0^2}\int^{2\pi}_0 d\theta
\int_0^\infty dq\, q\,  e^{-q^2+2a_0q\sin\theta}\nonumber\\
&\times& \sin(\Omega_qt)
\left(\begin{array}{c}
\cos(3\theta)+\eta\sin\theta \\ \sin(3\theta)-\eta\cos\theta \end{array}\right)
\end{eqnarray} 

and

\begin{eqnarray}\label{SpnExpt2}
\la S_z(t)\ra
&=&-\frac{3\hbar}{2\pi}
e^{-a_0^2}\int^{2\pi}_0 d\theta
\int_0^\infty dq\, q\,  e^{-q^2+2a_0q\sin\theta}\nonumber\\
&\times& \cos(\Omega_qt).
\end{eqnarray}

\begin{figure}[!]
\begin{minipage}[b]{\linewidth}
\centering
 \includegraphics[width=7.5cm, height=2.6cm]{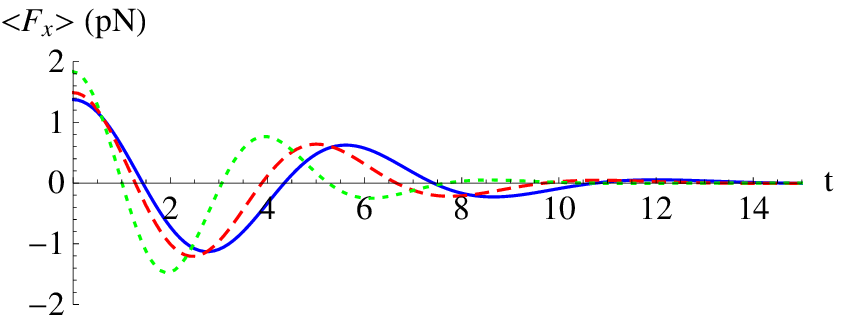}
\end{minipage}

\begin{minipage}[b]{\linewidth}
\centering
\includegraphics[width=7.5cm, height=2.6cm]{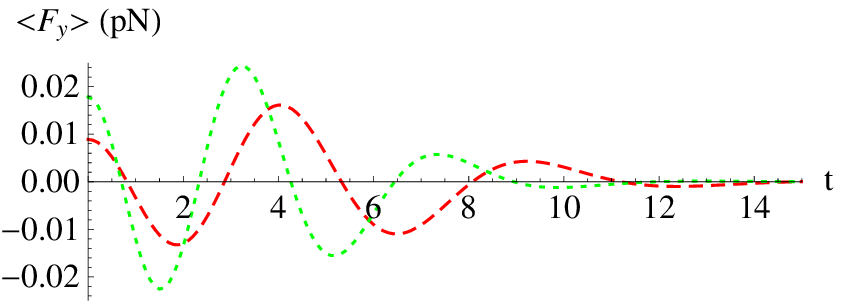}
\end{minipage}

\caption{Plots of the expectation values of spin-orbit force operators for $a_0=10$. Solid, dashed, and 
dotted line represent $\eta=0$, $0.5$, and $1$, respectively. Here, $t$ is considered in units of $10^{-3}\hbar d^3/(2\alpha)$.}
\label{ch3fig5}
\end{figure}

Let us now derive the expression of the spin-orbit force. 
The spin-orbit force is defined as ${\bf F}=m^\ast \ddot{\bf r}$, where $\ddot{\bf r}$
can be evaluated from Heisenberg equation $\ddot{\bf r}=-[[{\bf r},H],H]/\hbar^2$. Now,
it is straightforward to obtain ${\bf F}$ in the following operator form
\begin{eqnarray}
{\bf F}&=&\frac{2m^\ast p^2}{\hbar^7}\Big[(3\alpha^2+\beta^2)p^2-8\alpha \beta p_x p_y\Big]({\bf p}\times\hat{z})
\sigma_z.
\end{eqnarray}
Using Eq.~\eqref{ch3fr} we calculate the expectation values for the the components of
the spin force operator as
\begin{eqnarray}\label{forceExp}
\left(\begin{array}{c}
\la F_x(t)\ra \\\la F_y(t)\ra \end{array}\right)
&=&\frac{2m^\ast\alpha^2}{\pi\hbar^2d^5}
e^{-a_0^2}\int^{2\pi}_0 d\theta
\int_0^\infty dq\,  e^{-q^2+2a_0q\sin\theta}\nonumber\\
&\times& q^6\cos(\Omega_qt)g(\theta)\left(\begin{array}{c}
\sin\theta \\ -\cos\theta \end{array}\right).
\end{eqnarray} 

In Fig. \ref{ch3fig4} and Fig. \ref{ch3fig5}, we have shown the 
time dependence of the expectation values of spin and spin-orbit 
force operators. From Fig. \ref{ch3fig4} it is evident that $\la S_x(t)\ra$
vanishes in the absence of the Dresselhaus SOI. When $\beta=0$
the angular integrations in Eqs.~\eqref{SpnExpt1} and \eqref{SpnExpt2}
can be done exactly which gives a vanishing result for $\la S_x\ra$. The other 
components of the spin operator i.e. $\la S_y(t)\ra$ and $\la S_z(t)\ra$ 
can be found in the following forms
\begin{eqnarray}\label{spinYt}
\la S_y(t)\ra=3\hbar e^{-a_0^2}\int_0^\infty dq\,q e^{-q^2}I_3(2a_0q)\sin\Big(\frac{2\alpha}{\hbar d^3}q^3t\Big)
\end{eqnarray}

and 

\begin{eqnarray}\label{spinZt}
 \la S_z(t)\ra=3\hbar e^{-a_0^2}\int_0^\infty dq\,q e^{-q^2}I_0(2a_0q)\cos\Big(\frac{2\alpha}{\hbar d^3}q^3t\Big),
\end{eqnarray}
where $I_n(z)$ is the modified Bessel function of first kind.
Eqs.~\eqref{spinYt} and \eqref{spinZt} indicate that 
the spin precesses in the $yz$ plane.
For $\beta=0$, one can find that $\la F_y(t)\ra=0$. In this case, $\la F_x(t)\ra$ takes 
the following form
\begin{eqnarray}\label{ForceX}
\la F_x(t)\ra=F_0 e^{-a_0^2}\int_0^\infty dq\,q^6 e^{-q^2}I_1(2a_0q)
\cos\Big(\frac{2\alpha}{\hbar d^3}q^3t\Big),
\end{eqnarray}
where $F_0=12m^\ast\alpha^2/(\hbar^2 d^5)$.
Analyzing Eqs.~\eqref{spinYt}-\eqref{ForceX}, one can conclude that
the spin precession in the $yz$ plane essentially generates a transverse
spin-orbit force along $x$ direction. This fact is depicted in Fig. \ref{ch3fig5}. 

The scenario
changes significantly in the presence of a finite $\beta$. In this case the
angular integrations in Eqs.~\eqref{SpnExpt1}-\eqref{SpnExpt2} and Eq.~\eqref{forceExp}
can not be done exactly.
A numerical calculation shows that $\la S_x(t)\ra$
survives for $\beta\neq0$ but its magnitude is small compared to that of $\la S_y(t)\ra$
and $\la S_z(t)\ra$. Thus the spin starts to precess in the entire space as one switches on $\beta$.
As a result, a finite $y$ component of the spin-orbit force is generated. But the
amplitude of $\la F_y\ra$ is negligibly small in comparison to that of $\la F_x\ra$ as 
evident from Fig. \ref{ch3fig5}. So, one can safely consider that the spin-orbit
force acts along the $x$ direction provided the initial wave packet is injected along
the $y$ direction. In recent past\cite{zb2d1}, the origin of ZB was explained in the light of the spin-orbit 
force and it was argued that ZB occurs in a direction along which the spin-orbit force
acts on the wave packet. By observing Eqs.~\eqref{Ch3posX}, \eqref{ch3velX}, and \eqref{ForceX},
we conclude that our results are consistent with that prediction. 

\section{Persistent Zitterbewegung and Possible detection}
The ZB, manifested in the expectation values of the physical observables 
undergo transient behavior. However, the persistent behavior of ZB can be
obtained in the limit $a_0\gg1$ i. e. the initial wave packet is sharply peaked in the
momentum space. For $\beta=0$, one can obtain the following approximate
expressions for the expectation values:
$\la x\ra\simeq(3/2k_0)[1-\cos(\omega_0t)]$, $\la v_x\ra\simeq(3\alpha k_0^2/\hbar)\sin(\omega_0t)$, 
$\la S_y\ra\simeq(3\hbar/2)\sin(\omega_0t)$, $\la S_z\ra\simeq(3\hbar/2)\cos(\omega_0t)$, and 
$\la F_x\ra\simeq f_0\cos(\omega_0t)$ where $\omega_0=2\alpha k_0^3/\hbar$ and $f_0=6m^\ast\alpha^2k_0^5/\hbar^2$.
Due to the presence of SOI, an
effective Lorentz-like magnetic field\cite{pal} can be obtained in the following form
${\bf B}_{\rm s}=2m^{\ast 2}\alpha^2p^4g(\theta)\sigma_z\hat{z}/(e\hbar^7)$.
In the $a_0\gg1$ limit and when $\beta=0$ we obtain 
$\la B_{\rm s}\ra \simeq B_0\cos(\omega_0t)$ where $B_0=6m^{\ast2}\alpha^2k_0^4/(e\hbar^3)$.
Interestingly, we find
$\la F_x\ra/\la B_{\rm s}\ra=e\hbar k_0/m^\ast$  
i.e. $F_x$ plays a role of the Lorentz force in the $a_0\gg1$ limit.
Now, the time dependent magnetic field $B_s$ would generate an electric voltage
of the following form $V=V_0\sin(\omega_0t)$ where $V_0=\omega_0AB_0$ with
$A$ as the area of the sample. For typical material parameters a GaAs quantum well
i.e. $m^\ast=0.45m_e$, $\alpha=10^{-28}$ eVm$^3$, $A=10^4$ nm$^2$, and $k_0=10^8$ m$^{-1}$ we obtain
$\omega_0\simeq 3\times 10^{11}$ rad/s, $B_0\simeq 0.14$ T, and $V_0\simeq 4.4\times10^{-4}$ V.
In principle, one can measure $V_0$ to confirm the signature of ZB experimentally.
As introduced recently\cite{zb2d5} this kind of electric voltage associated with ZB can be also obtained
in the case of a 2DEG with time dependent linear Rashba SOI.

\section{Summary}
In summary, we have investigated the $zitterbewegung$ of a wave packet in a 
2DHG in the presence of both Rashba and Dresselhaus SOIs. The $zitterbewegung$ manifested 
in position, velocity, and spin appear as transient oscillations. For equal strength of 
Rashba and Dresselhaus SOIs we find $zitterbewegung$ does not vanish, a direct
contradiction to the electronic case. This non-vanishing of $zitterbewegung$ has been 
interpreted as a consequence of the Berry phase. 
The pseudo-spin associated with the heavy hole precesses about a SOI induced effective magnetic field. As a consequence 
of this spin precession, a transverse component of the spin-orbit force appears which in turn 
induces $zitterbewegung$ in position and velocity. The transverse spin-orbit force also generates an electric 
voltage. We have concluded that an indirect signature of $zitterbewegung$ can be unveiled by measuring this
voltage.

\end{document}